\documentclass{aa}

\usepackage{graphicx}
\usepackage{natbib}
\usepackage{amssymb}
\usepackage{txfonts}
\usepackage{ulem}
\usepackage[figuresright]{rotating}
\bibpunct{(}{)}{;}{a}{}{,} % to follow the A&A style

\newcommand{\ud}{\mathrm{d}} 
\newcommand{\reff}{\mbox{$R_\mathrm{eff}$}}
\newcommand{\msun}{\mbox{$M_{\odot}$}}

\newcommand{\bv}{\mbox{$B\!-\!V$}}
\newcommand{\vi}{\mbox{$V\!-\!I$}}

\newcommand{\bvzero}{\mbox{$(B\!-\!V)_{0}$}}
\newcommand{\vizero}{\mbox{$(V\!-\!I)_{0}$}}
\newcommand{\jhzero}{\mbox{$(J\!-\!H)_{0}$}}

\newcommand{\ebv}{\mbox{$E(B\!-\!V)$}}
\newcommand{\evi}{\mbox{$E(V\!-\!I)$}}
\newcommand{\hst}{\emph{HST}}
\newcommand{\acs}{\emph{ACS}}
\newcommand{\wfpc}{\emph{WFPC2}}
\newcommand{\nicmos}{\emph{NICMOS}}

\newcommand{\ishape}{{\sc Ishape}}

\newcommand{\iraf}{{\sc IRAF}}
\newcommand{\daophot}{{\sc DAOPHOT}}

\newcommand{\galev}{{\sc GALEV}}

\newcommand{\kms}{\ensuremath{\mathrm{km}\,\mathrm{s}^{-1}}}
\newcommand{\bband}{\emph{F435W}}
\newcommand{\vband}{\emph{F555W}}
\newcommand{\iband}{\emph{F814W}}
\newcommand{\jband}{\emph{F110W}}
\newcommand{\hband}{\emph{F160W}}

\begin{document}

\title{A peculiar object in M51: fuzzy star cluster or
  a background galaxy?\thanks{Based
on observations made with the NASA/ESA \emph{Hubble Space Telescope},
obtained from the data archive at the Space Telescope Institute. STScI
is operated by the association of Universities for Research in
Astronomy, Inc., under the NASA contract NAS 5-26555.}}

   \titlerunning{A peculiar object in M51}

   \author{R. A. Scheepmaker\inst{1} 
	\and
	H. J. G. L. M. Lamers\inst{1}
       \and
       S. S. Larsen\inst{1}
       \and
       P. Anders\inst{1}
}

   \authorrunning{R. A. Scheepmaker et al.}

\offprints{R. A. Scheepmaker, \email{scheepmaker@astro.uu.nl}}

\institute{$^1$Astronomical Institute, Utrecht University,
  Princetonplein 5, NL-3584 CC Utrecht, The Netherlands}

\date{Received 26 July 2007 / Accepted 19 October 2007}

%______________________________________________________________________ABSTRACT
\abstract
  % context heading (optional)
  % {} leave it empty if necessary (1st and last one) 
% Context (optional):
{}
% Aims:
{ We study a peculiar object with a projected position close to the
  nucleus of M51. It is unusually large for a star cluster in M51 and
  we therefore investigate the three most likely options to explain
  this object: (a) a background galaxy, (b) a cluster in the disk of M51 and (c) a
  cluster in M51, but in front of the disk.}
% Methods:
{ {\rm We use broad-band images of the Advanced Camera for Surveys and
the Near Infrared Camera and Multi-Object Spectrometer}, both on board
the \emph{Hubble Space Telescope}, to study the properties of this
object. Assuming the object is a star cluster, we fit the metallicity,
age, mass and extinction using simple stellar population
models. Assuming the object is a background galaxy, we estimate the
extinction from the colour of the background around the object. We
study the structural parameters of the object by fitting the spatial
profile with analytical models.}
% Results:
{ We find de-reddened colours of the object which are bluer than
  expected for a typical elliptical galaxy, and the central surface
  brightness is brighter than the typical surface brightness of a disc
  galaxy. It is therefore not likely that the object is a background
  galaxy. Assuming the object is a star cluster in the disc of M51, we
  estimate an age and mass of $0.7^{+0.1}_{-0.1}$~Gyr and
  $2.2^{+0.3}_{-0.3}\times 10^{5}~\msun$, respectively (with the
  extinction fixed to $\ebv = 0.2$). Considering the large size of the
  object, we argue that in this scenario we observe the cluster just
  prior to final dissolution. If we fit for the extinction as a free
  parameter, a younger age is allowed and the object is not close to
  final dissolution. Alternatively, the object could be a star cluster
  in M51, but in front of the disc, with an age of
  $1.4^{+0.5}_{-0.2}$~Gyr, mass $M = 1.7^{+0.8}_{-0.3}\times
  10^{5}~\msun$. Its effective radius is between $\sim$12--25~pc. This
  makes the object a ``fuzzy star cluster'', raising the issue of how an
  object of this age would end up outside the disc.}
% Conclusions (optional):
{}


\maketitle
%###############################################################################
%###############################################################################

%___________________________________________________________________INTRODUCTION
\section{Introduction}
  \label{sec:Introduction}
 
Recently, \citet{scheepmaker07} studied the effective (half-light)
radii of 1284 star clusters in the disc of M51. The largest cluster
candidate in their sample, carrying their ID number ``212995'', is
remarkable in several ways.

If the object is a cluster in the disk of M51, it is exceptionally large,
with an estimated effective radius of 21.6~pc, and located at a
galactocentric distance of 1~kpc in a region with many dust lanes.  A
cluster with this size and galactocentric distance is not expected to
survive for long in the spiral disc, where the disruptive effects of
the tidal field and other external perturbations (spiral arms, GMCs)
are large \citep{gieles05, gieles06c}. If the cluster candidate is
indeed in the disc, it should therefore be young and highly reddened to
account for its observed colours of $\bv = 0.75$ and $\vi =
1.05$. Using only 2 colours, however, \citet{scheepmaker07} could not
accurately determine an age and extinction.

Observations of HII regions in the disc of M51 show that the
metallicity of the gas is approximately solar to twice solar
\citep{diaz91, hill97}. If the object is in the disc of M51, we
would therefore expect its metallicity to be around solar. Outside of the
disc, the metallicity could be lower.

There is also the possibility that the object is a background galaxy.
The observed colours of the object are similar to the colours of
typical background galaxies, but this suggests that the reddening
through the disc of M51 is negligible. This would be quite remarkable
at a distance of 1~kpc from the nucleus, where the extinction is
expected to be high, as suggested by the presence of numerous dust
lanes. The hypothesis of a background galaxy therefore suggests a
chance alignment, such that the object is observed through a ``hole'' in
a region of the disc with many dust lanes.

In this work we present a more detailed study of the nature of
the object by extending the data with archival \hst/\nicmos\
observations in \jband\ and \hband, and by estimating the extinction
of the object. Having better constraints on the intrinsic spectral
energy distribution of the object, we compare the two hypotheses, of
the object being a star cluster or a background galaxy, in a
quantitative way.

%______________________________________________________OBSERVATIONS_AND_PHOTOMETRY
\section{Observations and photometry}
  \label{sec:Observations and photometry}

\subsection{Observations}
  \label{subsec:Observations}

We make use of the \hst/\acs\ dataset in the \bband, \vband\ and
\iband\ bands of M51 (NGC~5194). These data were taken as part of the
Hubble Heritage Project in January 2005 under proposal ID~10452 (PI:
S.~V.~W.~Beckwith). We also used \hst/\nicmos\ (NIC3) observations in
the \jband\ and \hband\ bands from the archives (proposal ID~7237, PI:
N.~Scoville). The
\acs\ dataset is described in more detail in \citet{mutchler05} and
\citet{scheepmaker07}. For the \nicmos\ dataset, we refer to
\citet{bastian05a}.

\begin{figure}
\centering
 \includegraphics[width=85mm]{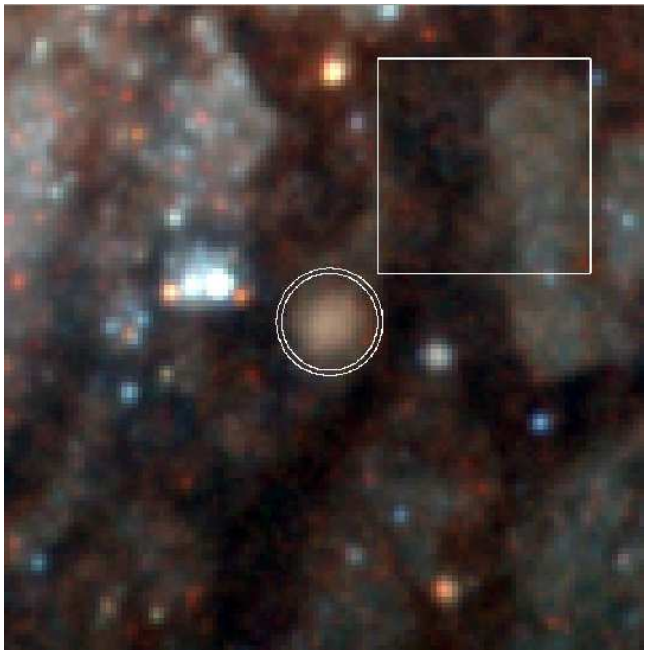}
\caption{Colour image of $120\times 120$ pixels centred on the
  peculiar object ``212995''. \bband\ is blue, \vband\ is green and
  \iband\ is red. The two rings have a radius of 9 and 10 pixels,
  respectively. The box shows the $40\times 40$ pixel region used for
  estimating the intrinsic colour of the background.}
\label{fig:RGB}
\end{figure}

\citet{scheepmaker07} located the peculiar object (carrying their ID
number 212995) at RA=13$^{\mathrm{h}}$29$^{\mathrm{m}}$51\fs94,
Dec=+47\degr11\arcmin19\farcs63. A colour image of a $120\times 120$
pixel region centred on the object is shown in Fig.~\ref{fig:RGB},
showing that the object is positioned in between two dark dust lanes.

%Figure rgb_100x100_9_10.ps, Created on SIU052 with 12995_makecolourzoom.pro

\citet{scheepmaker07} used the \acs\ data to perform aperture
photometry on the object. Their settings and aperture corrections
were, however, optimised for star clusters with an effective radius of
$\sim$3~pc.  We discuss new aperture photometry on the object in
Sect.~\ref{subsec:Photometry}, taking into account the larger size of
the object and its position in between the dust lanes.

The effective radius of 21.6~pc was measured by \citet{scheepmaker07},
who used the \ishape\ routine \citep{larsen99, larsen04b} to convolve
analytic surface brightness profiles with the point spread function
(PSF). EFF (or Moffat) profiles with a power-law index of $-1.5$ were
used (i.e. EFF15 profiles), because these were found to be the
best-fitting profiles for young stellar clusters in the LMC
\citep{elson87}. However, no tests were performed to find the
best-fitting surface brightness profile for our specific object. We
describe new fits to the surface brightness profile in
Sect.~\ref{sec:Surface brightness profile}, also using different
analytic profiles.

\subsection{Photometry}
  \label{subsec:Photometry}

We performed aperture photometry on the object using the
\iraf\footnote{The Image Reduction and Analysis Facility (\iraf) is
distributed by the National Optical Astronomy Observatories, which are
operated by the Association of Universities for Research in Astronomy,
Inc., under cooperative agreement with the National Science
Foundation.}/\daophot\ package. For the \acs\ data an aperture with a
radius of 8 pixels ($0.4\arcsec$) was used, and for the background
annulus we adopted an inner radius of 9 pixels and a width of 1
pixel. This annulus is indicated in Fig.~\ref{fig:RGB}. Zero points in
the VEGAmag system were taken from the \acs\ website.  There is a
possibility that the background annulus contains a small fraction of
light from the object. However, this will leave the colours of the
object, and therefore our results, largely unaffected, except for the
value of the mass (Sect.~\ref{subsec:Star cluster}), which will be a
lower limit.

For the \nicmos\ data we used an aperture with a radius of 2 pixels
($0.4\arcsec$) and a background annulus with an inner radius of 2
pixels and a width of 1 pixel. These values are consistent with the
aperture of the \acs\ photometry. The magnitudes in the VEGAmag system
were determined from the PHOTFNU values from the header, and the
bandpass averaged flux densities for Vega from the on-line photometric
keyword tables.

No aperture corrections were applied, since these are not well defined
for an extended object of which the nature is not well
known. However, the aperture was chosen to be as large as possible to
minimise any systematic errors. Not applying any
aperture correction will not change the colour of the objects
significantly (assuming there are no colour gradients), since
\citet{scheepmaker07} and \citet{bastian05a} have shown that the
aperture corrections are practically the same for the different
filters we use.

To facilitate reading, we will sometimes refer to the \hst\ filters
using the extended Johnson-Cousin notation (BVIJH), although we note
that \emph{no} transformations to this system have been applied. The
results of the photometry are listed in Table~\ref{tab:photometry}.

\begin{table}
\caption[]{Results of the aperture photometry, corrected for Galactic
  foreground extinction.}
\label{tab:photometry}
\begin{center}
\begin{tabular}{cc}
\hline\hline
Filter & Magnitude (VEGAmag) \\
\hline 
\bband\ ($\sim B$) & 22.13 \\
\vband\ ($\sim V$) & 21.50 \\
\iband\ ($\sim I$) & 20.49 \\
\hline
\jband\ ($\sim J$) & 20.08 \\
\hband\ ($\sim H$) & 19.26 \\
\hline
 \end{tabular}
\end{center}
\end{table}

%_____________________________________________________________________EXTINCTION
\section{Extinction}
  \label{sec:Extinction}

We have corrected the aperture photometry for Galactic foreground
extinction according to Appendix~B of \citet{schlegel98}. Since the
object is located in projection close to the nucleus of M51
($R_\mathrm{G} = 1.02$~kpc), there is possibly a significant amount of
additional extinction caused by the ISM of M51. We used two methods to
estimate this additional extinction of the object.  We note that from
here on, all quoted photometric values are already corrected for Galactic
extinction, and any extinction we will mention refers to the additional
extinction at the location of the object in M51.

\subsection{Extinction from cluster models}
  \label{subsec:Extinction from SSP models}

The first method used to estimate the extinction assumes that the object is
a star cluster, so we use cluster models to fit the extinction based
on broad-band colours.  We used the \galev\ simple stellar population
(SSP) models \citep{anders03} together with the {\sc AnalySED} code
\citep{anders04b} to simultaneously fit the extinction, metallicity,
age and mass of the cluster, based on our broad-band photometry in the
\bband, \vband, \iband, \jband\ and \hband\ bands. The
\citet{cardelli89} standard extinction law was used ($R_{\mathrm{V}} =
3.1$).  Table \ref{tab:galev} shows the results for different initial
constraints on the metallicity, which will be discussed in more detail
in Sect.~\ref{subsec:Star cluster}. These results are consistent with
$E(B-V) \leq 0.7$, with a best-fitting \ebv\ of 0.1. The resulting ages
and masses from these fits will be discussed in
Sect.~\ref{subsec:Star cluster}.

\subsection{Extinction from background colour}
  \label{subsec:Extinction from background colour}

The second method used to estimate the extinction of the object is based on
comparing the average colour of a background annulus around the object
to the intrinsic colour of the local background. The advantage of this
method is that we do not have to assume the nature of the object
(background galaxy or star cluster) a priori. However, the local
background might not be smooth, but could show intrinsic colour
variations. The intrinsic colour of the background is therefore not
well defined and this method can therefore only provide a rough
estimate of the extinction.

For the average observed colour of the background around the object we
used the same annulus as we used for the aperture photometry (i.e.\ an
inner radius of 9 pixels and a width of 1 pixel). After correcting for
Galactic foreground extinction, the ring has a mean \bv\ and \vi\ of
$0.75\pm 0.06$ and $1.29\pm 0.06$, respectively.

\begin{figure}
\centering
\includegraphics[width=85mm]{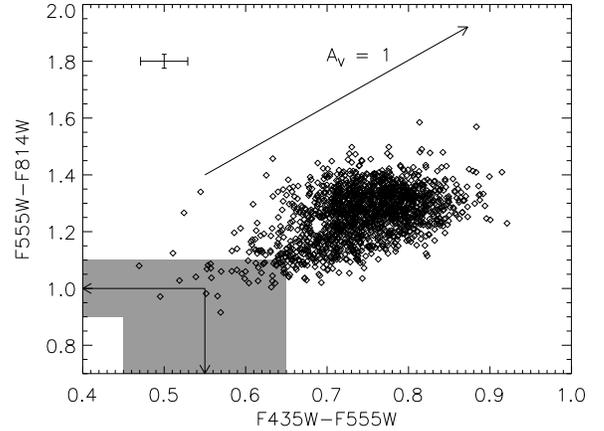}
\caption{$F435W-F555W$ versus $F555W-F814W$ diagram of the pixels in
  the background box, after correcting for Galactic foreground
  extinction. The extinction vector is the standard extinction law
  from \citet{cardelli89}. The cross in the top-left corner
  shows the typical error of the pixel colours. The gray region is selected by eye and
  shows the uncertainty around the two values selected as the
  intrinsic colour of the background (indicated by the two arrows):
  $\bvzero = 0.55\pm 0.10$, $\vizero = 1.00\pm 0.10$.}
\label{fig:pixels}
\end{figure}

To estimate the intrinsic colour of the background we used the $40
\times 40$ pixel region indicated with the box in
Fig.~\ref{fig:RGB}. This region was selected because it is free of
point sources and close to the object, and contains a large range of
pixel colours, which we assume to be mainly due to extinction
differences (the region is at the border of a dust lane). We show the
\bv\ versus \vi\ colour-colour diagram of all the pixels in this box
in Fig.~\ref{fig:pixels}, which also shows the extinction vector. The
observation that the scatter in this diagram is mostly parallel to the
extinction vector is consistent with extinction being the dominant
cause for colour variations among the pixels. Assuming that the pixels
with the lowest values for \bv\ and \vi\ are least extincted, we
estimate the upper limit for the intrinsic colour of the background to
be $\bvzero = 0.55\pm 0.10$ and $\vizero = 1.00\pm 0.10$.

We can estimate a lower limit for the extinction of the background
ring around the object by assuming that the ring has the intrinsic
colour of the background, and that the extinction is caused by a
foreground sheet. Thus, the observed mean colours of the ring result
in $\ebv = 0.2\pm 0.1$ and $\evi = 0.3\pm 0.1$. It is more likely,
however, that any dust present in the ring will be mixed with the
stars. For the same sheet of dust, this reduces the amount of
extinction estimated by the observed colours of the ring. We therefore
conclude that if our object is in the disc, it has an extinction of
\emph{at least} $\ebv = 0.2$.  If the object is behind the disc, the
extinction is likely to be even higher.

%Figure background_colourcolour.ps, Created with 212995_colourcolour.pro

%_____________________________________________________SURFACE_BRIGHTNESS_PROFILE
\section{Surface brightness profile}
 \label{sec:Surface brightness profile}

\begin{figure}
\centering
\includegraphics[width=85mm]{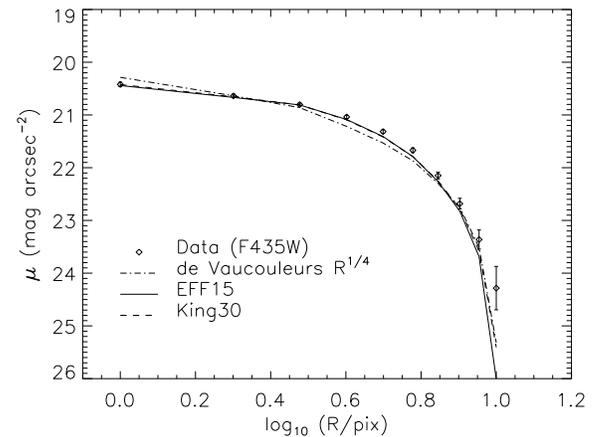}
\caption{Surface brightness profile of the object (including
    errorbars), compared to the best-fitting surface brightness
    profiles, assuming typical star cluster profiles (EFF15 and King
    30) and the typical profile for elliptical galaxies (a de
    Vaucouleurs $R^{1/4}$ law).}
\label{fig:profile}
\end{figure}

In order to study the structural parameters of our object, we measured
the surface brightness profile by performing aperture photometry in a
series of circular rings around the object, measuring the average
intensity in every ring as a function of radius. The resulting profile
is shown in Fig.~\ref{fig:profile}, in which we also plot the
best-fitting surface brightness profiles as determined by the \ishape\
routine \citep{larsen99, larsen04b}. \ishape\ fits two-dimensional
analytic profiles to the surface brightness distribution of a source,
taking into account the convolution with the PSF of the telescope. In
Fig.~\ref{fig:profile} we show the best-fitting profiles typically
used for fitting star clusters (i.e. a King~30 profile, see
\citealt{king62}, and an EFF15 profile, see \citealt{elson87}) and the
profile typically used for fitting elliptical galaxies (a de
Vaucouleurs $R^{1/4}$ profile, see \citealt{devaucouleurs48}). The
surface brightness profiles of these best-fitting models were measured
from the model images that \ishape\ creates. Therefore they include
the convolution with the PSF and take into account the slight
ellipticity of the object.

%Figure surfacebrightness_profile.ps, Created with paper_radialprofile.pro

Assuming different analytic profiles, we find consistent results for
the FWHM and the aspect ratio ($a$) of the object, namely ${\rm FWHM}
\approx 9$~pixels (along the major axis) and $a \approx
0.82$. Although the cluster profiles (King/EFF) provide better fits
than a de Vaucouleurs profile, \ishape\ can not determine the
best-fitting concentration/power-law index for these profiles, since
the extended ``wings'' are affected by the dust lanes too much. The
value for the best-fitting effective radius depends on this index, and
therefore on the choice of the profile. In Table~\ref{tab:radius} we
summarise the results of fitting the surface brightness distribution
of our object, assuming different profiles. For the sake of
completeness we also included the values for the de Vaucouleurs
profile. Due to the much steeper decline in the innermost regions of a
de Vaucouleurs profile, the FWHM of this profile is much smaller than
the FWHM of the other profiles, and the effective radius is much
larger. These values are not very meaningful, however, since the de
Vaucouleurs profile does not provide a good fit to the data.

For comparison, the effective radius of a typical star cluster is 2--3
pc \citep{scheepmaker07, jordan05, harris96}, which, at the distance
of M51, is equivalent to a ${\rm FWHM} \approx 1$ pixel. Thus, although
the effective radius of the object is not well constrained without
knowing the underlying profile, it is clear that if the object is at
the distance of M51, it is about 9--10 times larger than a typical
star cluster.

% Created with values from ishape_212995.bl 
\begin{table}
\caption[]{Results of fitting different analytic profiles to the
  observed surface brightness distribution on the \bband\ image, using the \ishape\ routine.}
\label{tab:radius}
\begin{center}
\begin{tabular}{lccc}
\hline\hline
Profile & FWHM (pix) & \reff\ (pix) & \reff\ (pc)$^{\mathrm{a}}$ \\
\hline 
de Vaucouleurs & 0.15 & $1.1\times 10^{3}$ & -- \\
EFF15 & 9.21 & 9.47 & 19.29$^{\mathrm{b}}$ \\
EFF25 & 9.34 & 5.78 & 11.77 \\
King 5 & 8.80 & 5.66 & 11.52 \\
King 30 & 9.07 & 12.15 & 24.74 \\
\hline
\end{tabular}
\begin{list}{}{}
\item[$^{\mathrm{a}}$]Assuming a distance to M51 of $8.4\pm 0.6$~Mpc
  \citep{feldmeier97}.
\item[$^{\mathrm{b}}$]This value is slightly smaller than the $\reff
  = 21.6$~pc measured by \citet{scheepmaker07}, which is caused by the
  use of a larger ``fitting radius'' in the current study (i.e. 10 pixels). 
\end{list}
\end{center}
\end{table}

\section{The nature of the object}
 \label{sec:The nature of the object}

%_____________________________________________________________BACKGROUND_GALAXY
\subsection{Background galaxy}
  \label{subsec:Background galaxy}

If the object is a background galaxy, the mean extinction of the ring
around the object (Sect.~\ref{subsec:Extinction from background
colour}) is a lower limit for the mean extinction of the background
galaxy. If we use an extinction of at least $\ebv = 0.2$, as
estimated in Sect.~\ref{subsec:Extinction from background colour}, to
correct the observed colours, we find intrinsic colours of the
background galaxy no redder than $\bvzero = 0.43$, $\vizero = 0.68$
and $\jhzero = 0.81$. In Table~\ref{tab:colours} we list
the corrected colours for a range of extinctions.

% Created with values from 212995_extinction_correction.pro 
\begin{table}
\caption[]{Intrinsic broad-band colours of the object, assuming
  different degrees of extinction and using the Galactic extinction
  law of \citet{cardelli89}.}
\label{tab:colours}
\begin{center}
\begin{tabular}{cccc}
\hline\hline
E(B-V) & $(B-V)_{0}$ & $(V-I)_{0}$ & $(J-H)_{0}$ \\
\hline 
0 & 0.63 & 1.00 & 0.82 \\
0.1 & 0.53 & 0.84 & 0.79 \\
{\bf 0.2$^{\mathrm{a}}$} & {\bf 0.43} & {\bf 0.68} & {\bf 0.76} \\
0.4 & 0.23 & 0.35 & 0.71 \\
0.6 & 0.03 & 0.03 & 0.65 \\
0.8 & -0.17 & -0.29 & 0.59 \\
1.0 & -0.37 & -0.62 & 0.54 \\
\hline
\end{tabular}
\begin{list}{}{}
\item[$^{\mathrm{a}}$] This value is our best estimate for the average
    extinction in a ring around the object
    (Sect.~\ref{subsec:Extinction from background colour}).
\end{list}
\end{center}
\end{table}

We can compare the intrinsic colours of the background galaxy to the
typical colours of other galaxies. \citet{tinsley71} finds $\bvzero
\approx 0.9$ for giant elliptical galaxies (after applying
K-corrections), and using the catalog of \citet{eisenhardt07}, we find
a mean $\bvzero = 1.15$ and $\vizero = 1.31$ for 486 galaxies in the
Coma cluster (without K-corrections). \citet{tonry01} find a mean
colour of $\vizero = 1.15\pm 0.06$ for a survey of 300 (mostly E and
S0) galaxies (without K-corrections). These colours are significantly
redder than those estimated for our object.  The colours used in these
galaxy surveys are in the standard Johnson-Cousins system, and a
direct comparison to the colours of the background galaxy could
therefore be misleading. However, \citet{fukugita95} give colours (not
corrected for Galactic reddening) of galaxies with different
morphological types in the \hst/\wfpc\ filter system. For E- and
S0-type galaxies, these colours are $(V_{\rm F555W}-I_{\rm
F814W}) = 1.27$ and 1.12, respectively. These colours are still
significantly redder than those estimated for our object, and this
suggests that the colour differences involved in the transformations
between the photometric systems are small.

Applying K-corrections to the objects in the catalogue of
\citet{eisenhardt07} or \citet{tonry01} would make their colours
bluer, but also the colours of our object would have to be corrected
for its K-correction, depending on its unknown distance. To reconcile
the difference in colours between our object and these catalogue
galaxies, the K-corrections of these galaxies have to be $\sim$0.6 mag
\emph{more} negative (in \bv\ and \vi) than those of our
object. Considering the distance to the Coma cluster of $\sim
85\pm10$~Mpc \citep{jensen99}, the expected K-corrections are
$\lesssim -0.1$~mag \citep{bicker04}.  Therefore we conclude that
K-corrections \emph{cannot} reduce the difference in colours between
our object and the typical galaxies.
 
\begin{figure}
\centering
 \includegraphics[width=85mm]{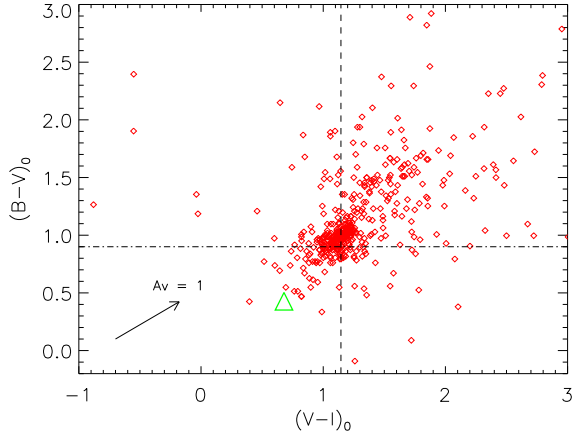}
\caption{\bvzero\ versus \vizero\ diagram, showing the 486 galaxies from the
  Coma cluster (red diamonds), the mean \bvzero\ of giant elliptical
  galaxies of \citet{tinsley71} (horizontal line) and the mean \vizero\
  of mostly E and S0 galaxies of \citet{tonry01} (vertical line). For
  comparison, we also show the estimated intrinsic colours of our
  object, assuming an extinction of $\ebv = 0.2$ (green triangle).}
\label{fig:colour-colour_galaxies}
\end{figure}

In Fig.~\ref{fig:colour-colour_galaxies} we compare the intrinsic
colours of the object to the typical galaxy colours.  If our object is
a background galaxy, its \bvzero\ and \vizero\ colours are
significantly bluer than the mean \bvzero\ and \vizero\ colours of
elliptical galaxies, and also bluer than 98\% and 97\%, respectively,
of the galaxies of the Coma cluster.

%Figure colour-colour.ps, Created with 212995_galaxies.pro

Another comparison to elliptical galaxies can be made in terms of the
surface brightness profile. The surface brightness profile of many
bright elliptical galaxies is best described by a de Vaucouleurs
$R^{1/4}$ law \citep{devaucouleurs48}.  In Fig.~\ref{fig:profile} we
compare the surface brightness profile of our object with a fit
assuming a de Vaucouleurs law. The figure shows that a de Vaucouleurs
law is steeper than the observed surface brightness profile (both an
EFF15 and a King 30 profile for clusters provide a better fit to the
data).

If our object is a background galaxy, it does not necessarily have to
be an elliptical or S0 galaxy, and could be a disc galaxy. For disc
galaxies, \citet{freeman70} found a nearly constant central surface
brightness of $\mu_{B} = 21.65\pm
0.30~\mathrm{mag}~\mathrm{arcsec}^{-2}$ for 28 out of 36 galaxies. 7
out of 36 galaxies had a surface brightness brighter than $\sim
20.1~\mathrm{mag}~\mathrm{arcsec}^{-2}$.  \citet{barden05} find
$\mu_{B} = 21.11\pm 0.03~\mathrm{mag}~\mathrm{arcsec}^{-2}$,
increasing with redshift to $\mu_{B} \approx 19.7\pm
0.07~\mathrm{mag}~\mathrm{arcsec}^{-2}$ for $z = 1$.  Comparing these
values to our object, within a radius of 2 pixels our object has
$\mu_{B} \leq 19.60~\mathrm{mag}~\mathrm{arcsec}^{-2}$, corrected for
Galactic foreground extinction, and an extinction of at least $\ebv
\geq 0.2$ for the disc of M51. Our object therefore has a surface
brightness which is brighter than the majority of the disc galaxies
studied by \citet{freeman70}, but it is consistent with the brightest
galaxies from this sample and with the highest redshift galaxies from
\citet{barden05}.

Assuming the lower-limit estimate of $\ebv = 0.2$, we can therefore not
exclude that the object is a background galaxy. We find, however, that
for an elliptical galaxy, the object has \bvzero\ and \vizero\ colours
$\sim 0.6$~mag bluer than expected and a surface brightness profile
that falls off less steeply than a $R^{1/4}$ law, while for a disc
galaxy the surface brightness is remarkably high.

%_____________________________________________________________STAR_CLUSTER
\subsection{Star cluster}
  \label{subsec:Star cluster}

Assuming the object is a star cluster, it could either be located in
the disc of M51, or in front of the disc and seen in projection.  For
the latter case, the metallicity of the cluster could be different
from the metallicity of the disc, and the extinction could be lower
than the mean extinction of the ring around the object
(Sect.~\ref{subsec:Extinction from background colour}).  We used the
\galev\ SSP models described in Sect.~\ref{subsec:Extinction from SSP
models} to fit the age, mass, extinction and metallicity of the
cluster. We should note here, however, that not using U-band data in
the SSP model fitting makes the derived ages and extinctions rather
uncertain. \citet{anders04b} have shown that not including U-band data
can lead to underestimating the extinction and overestimating the ages
of young clusters up to an order of magnitude. There are \hst/\wfpc\
U-band data of the region covering our object available in the
archives. However, these data are not deep enough to detect our
object.  We cannot exclude the possibility that our object is younger
than the best-fitting age determined from only using BVIJH. We will
take this uncertainty into account when we draw our conclusions in
Sect.~\ref{sec:Discussion and conclusions}.

The results of the fits, for different initial constraints, are
summarised in Table~\ref{tab:galev}.  Without any initial constraints
on the metallicity and extinction, the age and mass of the cluster are
not well defined, although the best fit indicates $Z/Z_{\odot} =
0.4$ and an extinction in the range $0$--$0.7$ in terms of \ebv. The
lowest two metallicities ($0.02Z_{\odot}$ and $0.2Z_{\odot}$) were
fitted with a lower probability than a metallicity in the range $0.4
\leq Z/Z_{\odot} \leq 2.5$.  We therefore exclude these lowest two
metallicities, which reduces the maximum possible age and mass of the
cluster to age~$\leq 2.1$~Gyr and $M \leq 2.8\times 10^{5} \msun$. A
metallicity in the range $0.4 \leq Z/Z_{\odot} \leq 2.5$ is consistent
with the observed metallicity in the inner disc of M51
(Sect.~\ref{sec:Introduction}).

% From values in /home/scheepmk/projects/m51/acs/212995/AnalySED/results.txt
\begin{table*}[!htbp]
\caption[]{Results of fitting different \galev\ SSP models to the
  broad-band photometry of the object.}
\label{tab:galev}
\begin{center}
\begin{tabular}{llccccc}
\hline\hline
\# & Constraint & $Z/Z_{\odot}$ & Age$/$Gyr &  $M/10^{5} \msun$ &
\ebv & Comment \\
\hline 
1. & none & 0.4 & $1.2^{+14.8}_{-1.2}$ & $1.6^{+7.5}_{-1.5}$ &
$0.1^{+0.6}_{-0.1}$ &  \\
2. &$ 0.4 \leq Z/Z_{\odot} \leq 2.5 $ & 0.4 & $1.2^{+0.9}_{-1.2}$ & $1.6^{+1.2}_{-1.5}$
   & $0.1^{+0.5}_{-0.1}$ & \\
3.& $Z = Z_{\odot}$ &  & $1.0^{+0.4}_{-1.0}$ & $1.9^{+0.6}_{-1.8}$ & $0.1^{+0.4}_{-0.1}$ & \\
4.& $Z/Z_{\odot} = 0.4$, $\ebv = 0$ & & $1.4^{+0.5}_{-0.2}$ & $1.7^{+0.8}_{-0.3}$ & & case B \\
5.& $Z = Z_{\odot}$, $\ebv = 0.2$ & & $0.7^{+0.1}_{-0.1}$ & $2.2^{+0.3}_{-0.3}$ & & case A \\
6.& $Z = Z_{\odot}$,  $\ebv = 0.7$ &  & $0.004^{\mathrm{a}}$ & $0.084^{\mathrm{a}}$ & & maximum extinction \\
\hline
\end{tabular}
\begin{list}{}{}
\item[$^{\mathrm{a}}$]The errors in this best-fitting model can not be estimated
  due to the much lower probability of the remaining models.    
\end{list}
\end{center}
\end{table*}

\begin{figure}
\centering
 \includegraphics[width=85mm]{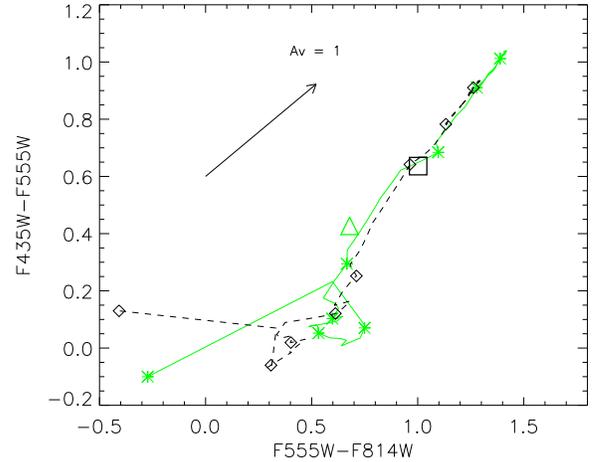}
\caption{$F435W-F555W$ versus $F555W-F814W$ diagram, showing the
  colour evolution with time of two \galev\ SSP models (green/solid:
  $Z = 0.02$, black/dashed: $Z = 0.008$ ), and the object without
  extinction correction (black square) and with an extinction
  correction of $\ebv = 0.2$ (green triangle). The symbols on the
  model curves indicate values of log(age), running from 6.6 (lower
  left) to 10.1 (top right) in steps of 0.5 dex.}
\label{fig:colour-colour_galev}
\end{figure}

If we fix the models to solar metallicity, we find an
extinction of $\ebv = 0.1^{+0.4}_{-0.1}$, which is roughly consistent with our
independent estimate of the extinction in the disc from
Sect.~\ref{sec:Extinction} (using the average background colour).  We
therefore consider two cases:
\begin{enumerate}
\item[A.] The cluster is located in the disc with an extinction of $\ebv \simeq
  0.2$ and solar metallicity.
\item[B.] The cluster is seen in projection against the disc with an
  extinction of $\ebv = 0.0$ and a metallicity of $Z/Z_{\odot} =
  0.4$.
\end{enumerate}
In Fig.~\ref{fig:colour-colour_galev} we compare the two cases with
the $F435W-F555W$ versus $F555W-F814W$ colour-colour evolution of
\galev\ SSP models for $Z/Z_{\odot} = 0.4$ (black dashed curve)
  and $Z/Z_{\odot} = 1$ (green solid curve).

%Figure colour-colour_galev_new.ps, Created with 212995_galev.pro

Considering case A, we find a best-fitting $\mathrm{age} =
0.7^{+0.1}_{-0.1}$~Gyr and $M = 2.2^{+0.3}_{-0.3}\times 10^{5} \msun$.
For case B we find an age of $1.4^{+0.5}_{-0.2}$~Gyr and a mass of $M
= 1.7^{+0.8}_{-0.3}\times 10^{5} \msun$.

If the cluster is located in the disk of M51 at a distance of only 1
kpc from the centre (case A), one must question how it can have
survived for such a long time. To this end we estimate its initial
mass and the expected disruption time.  For clusters losing mass due
to stellar evolution and tidal dissolution, the present mass is
related to the initial mass as
\begin{equation} \label{eq:mass}
M(t) = M_i \left\{ \mu_{\rm ev}(t)^{\gamma} - 
\frac{\gamma t}{t_0} M_i^{- \gamma} \right\}^{1/\gamma}
\end{equation}
with $M$ in \msun\ \citep{lamers05b}, where $\mu_{\rm ev}$ is the
fraction of the initial mass that would have remained at age $t$, if
stellar evolution had been the only mass loss mechanism.  For $t=0.7$
Gyr and solar metallicity, one finds $\mu_{\rm ev}=0.77$. The
parameters $t_0$ and $\gamma$ describe the mass loss due to
dissolution of the cluster as $(\ud M/\ud t)_{\rm dis}= -
M(t)^{1-\gamma}/t_0$ \citep{lamers05b}. $N$-body simulations of
clusters in tidal fields \citep{baumgardt03} and $N$-body
simulations of cluster dissolution due to shocks by encounters with
giant molecular clouds or spiral arms \citep{gieles06c, gieles07} both
suggest that $\gamma \simeq 0.62$. This value agrees with the value
that was derived empirically from cluster samples in different
galaxies \citep{boutloukos03, gieles05}. The value of the disruption
parameter $t_0$ depends strongly on the local conditions. For clusters
in M51 at $1 < R < 5$~kpc, \citet{gieles05} find a mean value of $t_0 =
6.6 \times 10^{5}$~yr.  Adopting these values, Eq.~\ref{eq:mass}
implies an initial mass of $M_i \simeq 4.5\times 10^{5}$~\msun, and we
find that the cluster has lost about 50\% of its initial mass.

In this estimate we have neglected the dependence of the dissolution
time with galactocentric distance and the large radius of the
cluster. The dissolution time of clusters is expected to decrease with
galactocentric distance as $t_0\propto R_G /V$, where $V$ is the
circular velocity at distance $R_G$ from the centre
\citep{baumgardt03, gieles05}. This implies that $t_0$ could be
smaller than the value adopted above by as much as a factor two or
three. In that case, Eq.~\ref{eq:mass} implies an initial mass of about 6--9$\times
10^{5}$~\msun, and the cluster has lost about 70\% of its mass.

We note that the cluster, having a \reff\ of $\sim$12--25 pc
 (depending on the assumed brightness profile, see
 Table~\ref{tab:radius}), is very extended and ``fuzzy''. The tidal
 radius of a $M = 2.2 \times 10^{5}$ \msun\ cluster at a
 galactocentric distance of 1.02 kpc is
\begin{equation}
r_{t} = \left(\frac{GM}{2V_{\mathrm{G}}^{2}}\right)^{1/3}R_{\mathrm{G}}^{2/3}
\approx 23.9~\mathrm{pc},
\end{equation}
where we used $V_{\mathrm{G}} = 200\kms$
\citep{garcia-burillo93}. Therefore, if the object is a cluster in the
disc of M51, its effective radius of $\sim$12--25 pc is very close to
the tidal radius (within a factor $\sim$0.5-1). This is remarkable,
since such an extended cluster would not be able to stay bound at this
location in the disc.

 The disruption time of a cluster depends implicitly on the radius
through the parameter $t_0$. For disruption in a tidal field, $t_0
\propto t_{\mathrm{rh}} \propto M^{1/2}r_{\mathrm{h}}^{3/2}$
\citep{spitzer87}, in which $t_{\mathrm{rh}}$ is the half-mass
relaxation time and $r_{\mathrm{h}}$ is the half-mass radius. This
suggests that more extended clusters have a longer disruption time
compared to compact clusters. However, for the disruption timescale
($t_{\mathrm{dis}}$) of clusters due to external perturbations
(e.g. shocks due to encounters with GMCs) \citet{spitzer58} has
derived that $t_{\mathrm{dis}} \propto M r_{\mathrm{h}}^{-3}$,
implying that a large cluster is more susceptible to external
disruptive effects (see also \citealt{gieles06c}). Considering the
location of the cluster close to the centre of M51 in a region with
many dust lanes, the external perturbations are likely to dominate in
the disruption, and they will shorten the disruption time for such an
extended cluster, compared to the disruption time of a cluster with a
typical size of about 3~pc. Therefore, the estimated value of the
disruption time parameter $t_0 \sim$~2--3$\times 10^{5}$~yr would be
too large. Using Eq.~\ref{eq:mass}, this implies that the initial mass
of such an extended cluster is probably much higher, to explain its
current mass and age. This could mean that the cluster has lost
significantly more than 70\% of its initial mass, and is currently
uncomfortably close to final dissolution. Although such a scenario is
not impossible, it is unlikely that we observe a cluster in this last
phase of its lifetime.

Case B, seeing the cluster in projection against the disc, is
considered to be more likely. In this scenario the cluster can be
located well outside the disc, where the tidal radius will be larger
and the expected disruption time will be longer. This means that the
initial mass needs to be lower compared to case A, and that we
observe the cluster in a more ``calm'' phase of its lifetime.

To establish the nature of the object with greater certainty than
presented in this study requires spectroscopic observations. These
will be challenging, however, considering the low brightness of the
object (Table~\ref{tab:photometry}). Deep U-band imaging, obtained in
good seeing conditions, would also help establish the age and
extinction with greater certainty.

%_____________________________________________________________
\section{Conclusions}
  \label{sec:Discussion and conclusions}

We have studied the photometry, extinction and structural parameters
of a peculiar object in M51 using \hst/\acs\ and \hst/\nicmos\
data. We have considered different possible scenarios for the object,
from which we conclude the following:
\begin{enumerate}
\item If we assume that the object is a background galaxy, we estimate
  that the extinction through the disc of M51 is at least $\ebv =
  0.2$. Correcting for this \emph{minimum} extinction, we find
  intrinsic colours of $\bvzero = 0.43$, $\vizero = 0.68$ and $\jhzero
  = 0.76$. These colours suggest that the object is not a background
  galaxy, because \bvzero\ and \vizero\ are $\sim$0.6~mag bluer than
  expected for a typical elliptical galaxy
  (Fig.~\ref{fig:colour-colour_galaxies}) and the surface brightness
  is $\sim2~\mathrm{mag}~\mathrm{arcsec}^{-2}$ brighter than a
  typical disc galaxy. If the extinction would be larger than $\ebv =
  0.2$, these differences in colour and surface brightness would be
  even larger.

\item The surface brightness distribution can be well approximated
  by typical cluster profiles (i.e.\ EFF or King profiles) with a
  FWHM of $\sim$9 pixels. A de Vaucouleurs $R^{1/4}$ law, typical for
  elliptical galaxies, does not provide a good fit to the inner part
  of the observed profile (Fig.~\ref{fig:profile}). Because the exact
  concentration (or power-law) index of the object can not be
  determined, the effective radius of the object depends on the choice
  of the profile. Assuming the object is at the distance of M51, our
  best estimates for the effective radius are between $\sim$12~pc and
  $\sim$25~pc (Table~\ref{tab:radius}).

\item If we assume that the object is a star cluster in the disc of
  M51, we estimate an age of $0.7^{+0.1}_{-0.1}$~Gyr and a mass $M =
  2.2^{+0.3}_{-0.3}\times 10^{5}~\msun$, assuming $\ebv = 0.2$ and
  solar metallicity. Considering the large size of the object,
  however, we estimate that in this scenario the cluster must
  initially have been very massive ($>4.5\times 10^{5} \msun$), and
  that we observe the cluster just prior to final
  dissolution. Although not impossible, this scenario is considered to
  be unlikely.

\item We can not fully exclude the possibility that our best-fitting
  cluster ages and extinctions are over- and underestimated,
  respectively, due to the lack of U-band photometry of our object. If
  we consider the \emph{maximum}-fitted extinction of $\ebv = 0.7$
  (model 6 in Table~\ref{tab:galev}), instead of the
  \emph{best}-fitted extinction, the cluster has an age of only
  $\sim$4~Myr and is not close to final dissolution. With our current
  data, however, the likelihood of this fit is very low.

\item If we assume that the object is a star cluster well outside the
  disc of M51 and seen in projection against it, we estimate an age of
  $1.4^{+0.5}_{-0.2}$~Gyr and a mass of $M = 1.7^{+0.8}_{-0.3}\times
  10^{5}~\msun$. This scenario naturally explains why the best-fitting
  extinction is low, $\ebv = 0.1^{+0.5}_{-0.1}$, although we observe
  the cluster in between two dust lanes. Furthermore, this scenario
  implies that we are not observing the cluster in a special phase of
  its lifetime.

\item We conclude that, without additional U-band data or
  spectroscopy, both the ``disc'' and ``projection'' scenarios are
  possible within the uncertainties. With the current data, however,
  the cluster models favour the scenario in which the object is a star
  cluster seen in projection against the disc. In this scenario, the
  cluster just qualifies as being a ``faint fuzzy'' star cluster,
  namely having $\bvzero > 0.6$, $\vizero > 1.0$ and $\reff > 7$~pc
  \citep{larsen00a}. Seen in projection against the spiral disc, the
  (projected) galactocentric distance of 1~kpc is not
  unexpected. However, with an estimated age of
  $1.4^{+0.5}_{-0.2}$~Gyr, the formation mechanism of such a cluster,
  in combination with its extended shape and its current location
  outside of the spiral disc, are interesting issues that leave room
  for further investigation.
\end{enumerate}

\noindent \emph{Note added in proof:} After the paper was accepted we
became aware of the study of background galaxies seen though the disk
of M~51 by \citet{holwerda05b}. These authors found that the average
opacity values for M~51 are the highest for any disk ($A_I = 1.5$),
making it very unlikely that the object is a background galaxy.

\begin{acknowledgements}
We thank the \emph{International Space Science Institute} (ISSI) in
Bern, Switzerland, where part of this research took place. The
anonymous referee is thanked for useful comments and suggestions.
\end{acknowledgements}

\bibliographystyle{aa} 
\bibliography{8359v2}

\begin{thebibliography}{36}
\expandafter\ifx\csname natexlab\endcsname\relax\def\natexlab#1{#1}\fi

\bibitem[{{Anders} {et~al.}(2004){Anders}, {Bissantz}, {Fritze-v.~Alvensleben},
  \& {de Grijs}}]{anders04b}
{Anders}, P., {Bissantz}, N., {Fritze-v.~Alvensleben}, U., \& {de Grijs}, R.
  2004, \mnras, 347, 196

\bibitem[{Anders \& Fritze-v. Alvensleven(2003)}]{anders03}
Anders, P. \& Fritze-v. Alvensleven, U. 2003, A\&A, 401, 1063

\bibitem[{{Barden} {et~al.}(2005){Barden}, {Rix}, {Somerville}, {Bell},
  {H{\"a}u{\ss}ler}, {Peng}, {Borch}, {Beckwith}, {Caldwell}, {Heymans},
  {Jahnke}, {Jogee}, {McIntosh}, {Meisenheimer}, {S{\'a}nchez}, {Wisotzki}, \&
  {Wolf}}]{barden05}
{Barden}, M., {Rix}, H.-W., {Somerville}, R.~S., {et~al.} 2005, \apj, 635, 959

\bibitem[{Bastian {et~al.}(2005)Bastian, Gieles, Lamers, Scheepmaker, \&
  de~Grijs}]{bastian05a}
Bastian, N., Gieles, M., Lamers, H. J. G. L.~M., Scheepmaker, R.~A., \&
  de~Grijs, R. 2005, A\&A, 431, 905

\bibitem[{Baumgardt \& Makino(2003)}]{baumgardt03}
Baumgardt, H. \& Makino, J. 2003, MNRAS, 340, 227

\bibitem[{{Bicker} {et~al.}(2004){Bicker}, {Fritze-v.~Alvensleben},
  {M{\"o}ller}, \& {Fricke}}]{bicker04}
{Bicker}, J., {Fritze-v.~Alvensleben}, U., {M{\"o}ller}, C.~S., \& {Fricke},
  K.~J. 2004, \aap, 413, 37

\bibitem[{Boutloukos \& Lamers(2003)}]{boutloukos03}
Boutloukos, S.~G. \& Lamers, H. J. G. L.~M. 2003, MNRAS, 338, 717

\bibitem[{{Cardelli} {et~al.}(1989){Cardelli}, {Clayton}, \&
  {Mathis}}]{cardelli89}
{Cardelli}, J.~A., {Clayton}, G.~C., \& {Mathis}, J.~S. 1989, \apj, 345, 245

\bibitem[{{de Vaucouleurs}(1948)}]{devaucouleurs48}
{de Vaucouleurs}, G. 1948, Annales d'Astrophysique, 11, 247

\bibitem[{{Diaz} {et~al.}(1991){Diaz}, {Terlevich}, {Vilchez}, {Pagel}, \&
  {Edmunds}}]{diaz91}
{Diaz}, A.~I., {Terlevich}, E., {Vilchez}, J.~M., {Pagel}, B.~E.~J., \&
  {Edmunds}, M.~G. 1991, \mnras, 253, 245

\bibitem[{{Eisenhardt} {et~al.}(2007){Eisenhardt}, {De Propris}, {Gonzalez},
  {Stanford}, {Wang}, \& {Dickinson}}]{eisenhardt07}
{Eisenhardt}, P.~R., {De Propris}, R., {Gonzalez}, A.~H., {et~al.} 2007, \apjs,
  169, 225

\bibitem[{Elson {et~al.}(1987)Elson, Fall, \& Freeman}]{elson87}
Elson, R. A.~W., Fall, S.~M., \& Freeman, K.~C. 1987, ApJ, 323, 54

\bibitem[{Feldmeier {et~al.}(1997)Feldmeier, Ciardullo, \&
  Jacoby}]{feldmeier97}
Feldmeier, J.~J., Ciardullo, R., \& Jacoby, G.~H. 1997, ApJ, 479, 231

\bibitem[{{Freeman}(1970)}]{freeman70}
{Freeman}, K.~C. 1970, \apj, 160, 811

\bibitem[{Fukugita {et~al.}(1995)Fukugita, Shimasaku, \& Ichikawa}]{fukugita95}
Fukugita, M., Shimasaku, K., \& Ichikawa, T. 1995, PASP, 107, 945

\bibitem[{Garc\'ia-Burillo {et~al.}(1993)Garc\'ia-Burillo, Combes, \&
  Gerin}]{garcia-burillo93}
Garc\'ia-Burillo, S., Combes, F., \& Gerin, M. 1993, A\&A, 274, 148

\bibitem[{{Gieles} {et~al.}(2007){Gieles}, {Athanassoula}, \& {Portegies
  Zwart}}]{gieles07}
{Gieles}, M., {Athanassoula}, E., \& {Portegies Zwart}, S.~F. 2007, \mnras,
  376, 809

\bibitem[{Gieles {et~al.}(2005)Gieles, Bastian, Lamers, \& Mout}]{gieles05}
Gieles, M., Bastian, N., Lamers, H. J. G. L.~M., \& Mout, J.~N. 2005, A\&A,
  441, 949

\bibitem[{{Gieles} {et~al.}(2006){Gieles}, {Portegies Zwart}, {Baumgardt},
  {Athanassoula}, {Lamers}, {Sipior}, \& {Leenaarts}}]{gieles06c}
{Gieles}, M., {Portegies Zwart}, S.~F., {Baumgardt}, H., {et~al.} 2006, \mnras,
  371, 793

\bibitem[{Harris(1996)}]{harris96}
Harris, W.~E. 1996, AJ, 112, 1487

\bibitem[{{Hill} {et~al.}(1997){Hill}, {Waller}, {Cornett}, {Bohlin}, {Cheng},
  {Neff}, {O'Connell}, {Roberts}, {Smith}, {Hintzen}, {Smith}, \&
  {Stecher}}]{hill97}
{Hill}, J.~K., {Waller}, W.~H., {Cornett}, R.~H., {et~al.} 1997, \apj, 477, 673

\bibitem[{{Holwerda} {et~al.}(2005){Holwerda}, {Gonzalez}, {Allen}, \& {van der
  Kruit}}]{holwerda05b}
{Holwerda}, B.~W., {Gonzalez}, R.~A., {Allen}, R.~J., \& {van der Kruit}, P.~C.
  2005, \aj, 129, 1396

\bibitem[{{Jensen} {et~al.}(1999){Jensen}, {Tonry}, \& {Luppino}}]{jensen99}
{Jensen}, J.~B., {Tonry}, J.~L., \& {Luppino}, G.~A. 1999, \apj, 510, 71

\bibitem[{Jord\'an {et~al.}(2005)Jord\'an, C\^ot\'e, Blakeslee, Ferrarese,
  McLaughlin, Mei, Peng, Tonry, Merritt, Milosavljevi\'c, Sarazin, Sivakoff, \&
  West}]{jordan05}
Jord\'an, A., C\^ot\'e, P., Blakeslee, J.~P., {et~al.} 2005, ApJ, 634, 1002

\bibitem[{King(1962)}]{king62}
King, I. 1962, AJ, 67, 471

\bibitem[{Lamers {et~al.}(2005)Lamers, Gieles, Bastian, Baumgardt, Kharchenko,
  \& Portegies~Zwart}]{lamers05b}
Lamers, H. J. G. L.~M., Gieles, M., Bastian, N., {et~al.} 2005, A\&A, 441, 117

\bibitem[{Larsen(1999)}]{larsen99}
Larsen, S.~S. 1999, A\&A Suppl. Ser., 139, 393

\bibitem[{Larsen(2004)}]{larsen04b}
Larsen, S.~S. 2004, A\&A, 416, 537

\bibitem[{{Larsen} \& {Brodie}(2000)}]{larsen00a}
{Larsen}, S.~S. \& {Brodie}, J.~P. 2000, \aj, 120, 2938

\bibitem[{Mutchler {et~al.}(2005)Mutchler, Beckwith, Bond, Christian, Frattare,
  Hamilton, Hamilton, Levay, Noll, \& Royle}]{mutchler05}
Mutchler, M., Beckwith, S. V.~W., Bond, H.~E., {et~al.} 2005, BAAS, 37

\bibitem[{{Scheepmaker} {et~al.}(2007){Scheepmaker}, {Haas}, {Gieles},
  {Bastian}, {Larsen}, \& {Lamers}}]{scheepmaker07}
{Scheepmaker}, R.~A., {Haas}, M.~R., {Gieles}, M., {et~al.} 2007, \aap, 469,
  925

\bibitem[{Schlegel {et~al.}(1998)Schlegel, Finkbeiner, \& Davis}]{schlegel98}
Schlegel, D.~J., Finkbeiner, D.~P., \& Davis, M. 1998, ApJ, 500, 525

\bibitem[{Spitzer(1958)}]{spitzer58}
Spitzer, L. 1958, ApJ, 127, 17

\bibitem[{Spitzer(1987)}]{spitzer87}
Spitzer, L. 1987, Dynamical evolution of globular clusters, Princeton Series in
  Astrophysics (Princeton University Press)

\bibitem[{{Tinsley}(1971)}]{tinsley71}
{Tinsley}, B.~M. 1971, \apss, 12, 394

\bibitem[{{Tonry} {et~al.}(2001){Tonry}, {Dressler}, {Blakeslee}, {Ajhar},
  {Fletcher}, {Luppino}, {Metzger}, \& {Moore}}]{tonry01}
{Tonry}, J.~L., {Dressler}, A., {Blakeslee}, J.~P., {et~al.} 2001, \apj, 546,
  681

\end{thebibliography}

\end{document}